# Non-linear effects in TXRF spectroscopy: Study on the influence of the sample amount.

[a]Bennun, Leonardo; [b]Piol, María Natalia and [b]Vázquez, Cristina.

[a]Laboratorio de Física Aplicada, Departamento de Física, Facultad de Ciencias Físicas y Matemáticas, Universidad de Concepción, Concepción, Chile.

[b]Universidad de Buenos Aires, Facultad de Ingeniería, Instituto de Química Aplicada a la Ingeniería, Laboratorio de Química de Sistemas Heterogéneos, Av. Paseo Colón 850, C1063ACV, Buenos Aires, Argentina.

Corresponding Author: L. B. E-mail: lbennun@udec.cl

**Abstract.**

In this work we have studied the limitations of the TXRF spectroscopy in the upper limit of validity of the technique, when the analyzed specimen ceases to be a thin film. We have evaluated the non-linear effects in spectra obtained from samples made "ad-hoc", which were acquired in a pre-established sequence of dilutions. So, the spectra were obtained in conditions of high, medium and low counts, thus the nonlinear effects were sequentially evaluated, as a function of the sample amount. According to the conclusions of this work, the Dead Time should be avoided since its effects are unpredictable and irreversible once the spectra are obtained.

## 1. Introduction

Total reflection X-ray fluorescence (TXRF) is a mature multi-elemental technique of analysis, often used to determine elemental composition of particles, residues, and impurities on smooth surfaces. TXRF is an energy dispersive XRF technique arranged in a special geometry. An incident X-ray beam impinges upon a polished flat sample carrier at very small angles, resulting in the reflection of most of the excitation beam photons at this surface. Due to this configuration, the measured spectral background in TXRF is less than in conventional XRF. This reduction results in a high increase of the signal to noise ratio. The dry specimen deposited in the sample carrier, is seen as a very thin layer, so the acquired spectrum is less prone to matrix influence (there is no need for corrections for attenuation or enhancement effects). Due to the elimination of matrix effects or in absence of limitations in the electronic data processing, that is, in proper conditions of the technique, the intensities of the TXRF peaks are directly proportional to the concentration of each element in the sample. This property gives the quantitative character to the technique.[1]

TXRF determines qualitatively and quantitatively elements between the atomic numbers 13 and 92. It is commonly accepted that the technique show a dynamic range of 5 orders of magnitude, with sensitivities between ng/mL in liquids and weight percent (% wt) in solids.[2]

Although the characteristics of TXRF, its uncertainties and, in general, the quality of its results have been extensively studied [3], e.g. by performing various ring tests [4], there is still one problem that has not yet been studied in detail: the upper limit of the sample quantity. Under this condition, the relationship between the concentration and the intensity of each element is no longer linear. The increase of the sample amount produces a loss of linearity, mainly for two reasons: 1) Due to physical interactions, from matrix effects, when the specimen cannot be considered as a thin film, and 2) due to limitations in the electronic response, produced by the Dead Time. As the rate of X-rays reaching the detector increases, it produces more signals to be processed by the detection chain. Since the electronic only can process a single event at the same time, the limitations of the response of the electronic system gradually begin to manifest in the processing of a large signal rate. Both limitations produce non-linear effects in the acquired spectrum, the first is related to the matrix effects and the second to the Dead Time.

In this work we have studied the limitations of the TXRF spectroscopy in the upper limit of validity of the technique, when the analyzed sample ceases to be a thin film. The spectra were obtained in conditions of high, medium and low counts, thus the nonlinear effects were sequentially evaluated. The activities in this work can be divided in 3 main stages: 1) The preparation and measuring of the samples, 2) Data analysis and interpretation, and 3) The application of the conclusions to unknown samples, in order to improve the reliability of the results, if it would be possible.

In all Sections in this work it is implicit that the experimentally acquired spectra are affected by statistical fluctuations. This characteristic also affects to the theoretical procedures and data analysis. In all cases we have neglected the influence of the statistical fluctuations, which are indeed second order phenomena in the cases studied here.

## 1.1. Technical details.

A proper quantification in TXRF analysis requires a linear relationship between the concentration of a given element, excited in the specimen, and its respective acquired peak intensity. This requirement is independent of the method of quantification used, the most common being the internal standard addition and the fundamental parameters method, and some variants can be found [5,6]. This condition is known as "thin film" since it is obtained when the deposited mass of the specimen is sufficiently small.

Usually, the thin film condition is considered as a kind of binary state, which distinguishes between an ideal situation or a non-acceptable one. If the specimen fulfills this requirement the measurement is considered as reliable, otherwise, it should be rejected, or at most only qualitative information can be obtained from the obtained spectrum. But insufficient current information can be found about this condition. Some questions, such as: What is the threshold (the limit) that defines the thin film approximation in TXRF? Which parameters describe how close or far the sample is to respect to the thin film condition? If the sample does not meet the thin film requirements, is there a function $G(E)$ that modifies the information of the obtained spectrum to increase the reliability of its interpretation? If so: 1) Which are the governing parameters and the validity limits of this function? 2) Would it be advisable to process the spectra with new programs instead of those usually used in TXRF? Many questions about the reliability of the obtained spectra need to be answered if the thin film condition is not assured/guaranteed.

In order to follow a systematic study of the non-linear effects in TXRF, first we describe the phenomena causing this non-desired response. There are two main sources: 1) From physical interactions of the excitation source over the atoms of the specimen, if the size of the specimen increases (such as matrix effects, which include enhancement and self-absorption, or scape peaks in the detector, etc.) and 2) From the instrumental characteristics of the used TXRF device (such as Dead Time, sum peaks (also known as pile-up), etc.). Although both phenomena are mixed, and their effects are intermingled in the obtained spectrum, in a first instance we are going to describe them separately.

### 1.2. Matrix Effects.

As the size of the sample increases: 1) The photons of the excitation source should travel an increasing distance inside the specimen in order to affect (excite) the inner atoms. The intensity of the excitation source is decreased as a function of the penetration length and is strongly dependent on the sample composition. Therefore, the inner atoms become progressively more invisible to this decreasing inhomogeneous excitation source. Moreover, if these inner atoms are excited, 2) their characteristic emission energies (X-ray photons) also should travel greater distances with attenuation properties on their way to reach the detector, and 3) the characteristic emissions of the sample may act as a secondary excitation source to the atoms in the same sample. This phenomenon is called "enhancement" and it is more pronounced with atoms with correlative atomic numbers.

### 1.3. Dead Time

If an X-ray photon arrives at the detector, a sequence of signals is triggered, until this information is properly acquired in the multi-channel analyzer. While this process is taken place, the system is unable to process another event; this instrumental limitation is called Dead Time. Let's suppose that a particular measuring system has a Dead Time of 20 µs, so the maximum number of events which it can process per second, is: $x = (1\ \text{s}/(20\ 10^{-6}\ \text{s})) = 50.000$ counts. If 100.000 photons per second are arriving at the detector, around the half will be irremediably lost. In this work we have studied how the absence of these non-collected photons, owing to the high counting rate, affects to the acquired spectrum. From a theoretical point of view, the statistical modeling of this effect is very complex, since there is a huge number of possible combinations. Being the Dead Time a phenomenon common to many spectroscopies (like Gamma Ray Spectroscopy used in Neutron Activation Analysis [7], alpha spectroscopy, etc.), in many of them (perhaps in most of them) the proper practice requires a non-high counting rate.

In order to carry out a study of the Dead Time, in many spectroscopies it is possible to modify the separation of the sample-detector system, in order to vary the geometric efficiency of detection. Therefore, the counting rate usually is controlled by this mechanism. Furthermore, in the classic detection chains for atomic or nuclear spectroscopies, the NIM modules allowed the simple change of different amplifiers, monochannel analyzers, amplitude to time converters (TACs), etc. In all of these analogue electronics, the user could make many changes in the settings, and also a complete calibration of the detection chain could be made, e.g. making pole-zero corrections. Any change in the instrumental settings may affect the Dead Time of the system.

In the case of TXRF instrumentation there was a strong change from analogue to digital electronic. The development of new technologies of X-ray sources and high resolution detectors, such as SDD (Silicon Drift Detector), have allowed the minimization of the equipment [8]. Now there are commercial compact TXRF equipment on the market, such as the S2 PicoFox or the S4 TStar (Bruker AXS, Germany) [9,10] or the Nano Hunter (Rigaku, Japan) [11]. The modern TXRF equipment in many ways are like computers, with solid-state modules, electronic boards, etc. Their repairs or upgrades are done by changing boards, like the changes of RAM, video cards, hard drives, etc. in a computer. In these modern TXRF devices there are very few external settings and details in the setup that can be modified, like the fine gain (amplification in the energy axis, $x$) or the adjustment of the angle of incidence of the excitation source on the sample holder. The rest of the settings are mainly selected by software, for instance, the procedure to adjust the shape or parameters of the peaks, (e.g. the FWHM), the method to determine the net counts of a particular line, (e.g. Least Square Marquardt fit procedure or Partial Least Squares (PLS) regression method, etc.), the automatic or manual search functions, the automatic calibration of energy, the subtraction of the background, etc.

For these reasons, perhaps the best way to study the Dead Time in modern TXRF spectroscopy is modifying the amount of the specimen deposited on the sample holder.

### 1.4. Data Analysis

In the analysis proposed here we do not make a complete taxonomy of the physical interactions that occur in TXRF spectroscopy. Instead, we only focus in the Photoelectric Interactions, which produce the useful information in the TXRF spectrum.

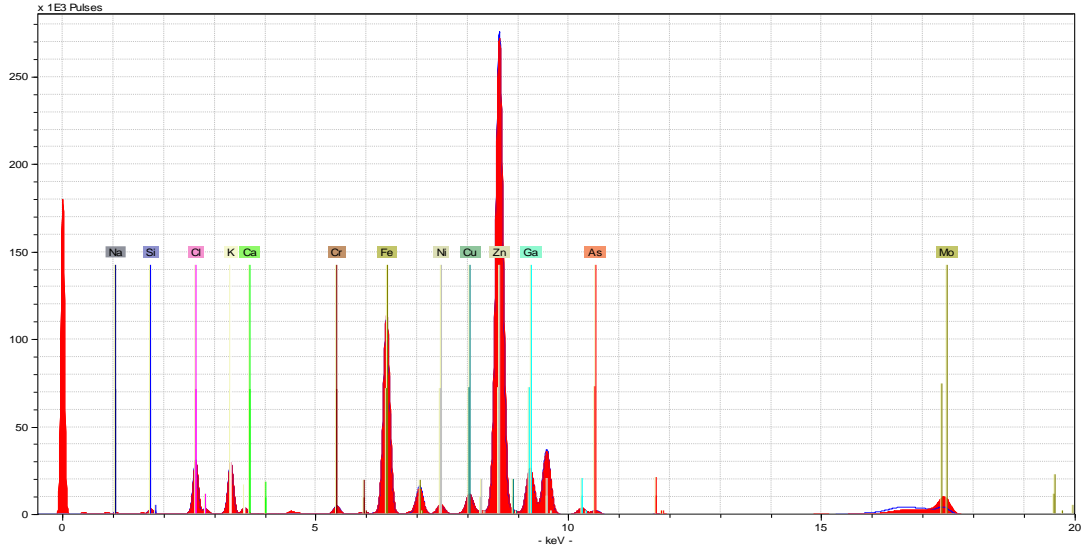

***Figure 1***: *Characteristic TXRF spectrum of the samples analyzed here, where the Photoelectric Interactions for the main elements evaluated in this work, can be observed.*

In this work we made a parametric study of the peak´s intensity, and on the shape of the complete spectrum, as a function of the amount of sample analyzed. The linear relationship between the number of counts of each peak (produced by Photoelectric Interactions), and the concentration of a given element, should be more reliable as the amount of the specimen reduces.

## 2. Theory

The basic equation for TXRF is:

$$I_i = I_0 N_0 K_i \frac{\sigma_i \omega_i m_i}{A_i} \qquad (1)$$

Where $I_i$ is the intensity of the signal produced by the element $i$, which is described in terms of the following parameters: $K_i$, the efficiency of detection, which depends mainly on geometric factors and on the energy of the line considered; $I_o$, is the intensity of the excitation source; $N_o$ the Avogadro´s Number; $m_i$, the mass surface density of element $i$ with atomic number Z and atomic mass $A_i$; $\sigma_i$, the excitation cross section; and $\omega_i$, the fluorescent yield.

Alternatively, the measured intensity ($I_i$) produced by the element $i$ in the spectrum can be expressed as:

$$I_i = I_0 S_i C_i \qquad (2)$$

Where $C_i$ is the element concentration in the analyzed sample, and $S_i$ corresponds to the sensitivity of the detection system for the selected element and a given X-ray line.

For quantification purposes, a common method consists in an internal standard addition to the original sample. For this particular element $j$, included in a known concentration in the sample, the Eq.(2) can be expressed as:

$$I_j = I_0 S_j C_j \quad (3)$$

In the data processing, the concentrations of the original elements are referred to the concentration of the included standard. In an acquired spectrum, the unknown concentration of the elements that compose the sample, can be obtained as:

$$C_i = \frac{I_i S_j C_j}{I_j S_i} = \frac{I_i S_{j/i} C_j}{I_j} \quad (4)$$

Where $S_{j/i}$ is the relative sensitivity of detection, of the standard $j$ with respect to element $i$. This is a known characteristic of the detection system. Since $I_i$ and $I_j$ are obtained from the spectrum, and the $C_j$ value is previously known, the concentration of the $i$ element can be obtained. We can rearrange Eq.(4) as:

$$\frac{I_i}{I_j} = \frac{C_i}{C_j S_{j/i}} = constant \quad (5)$$

If the technique is working in appropriate conditions, it can be noticed from Eq.(5) that the measuring of the same sample at different stages of dilutions should produce the same value of the relationship $\frac{I_i}{I_j}$. This is because the Dilution Factor affects $C_i$ and $C_j$ in the same way, and $S_{j/i}$ is a constant. So we can evaluate the relationship $\frac{I_i}{I_j}$ at different stages of dilutions in order to determine non-linear effects in the acquired spectrum, as the amount of sample analyzed is increased.

Following with the notation used in Eq. (1), it is convenient to describe the complete acquired spectrum as the product of two vectors:

$$S = I_0 N_0 (F_{z=1}, F_{z=2}, ., F_{z=n}) \begin{pmatrix} (Km\sigma\omega/A)_{z=1} \\ (Km\sigma\omega/A)_{z=2} \\ \vdots \\ (Km\sigma\omega/A)_{z=n} \end{pmatrix} \quad (6)$$

The first vector contains information about the characteristic functions $F_i$ for each element $i$.[12] The characteristic signals are functions obtained in an empirical manner, with the spectrometer and the use of a pure specimen of the element $i$ in question.[13] Each one is defined in fixed energy regions of the spectrum and we require them to be normalized, that is, the integral of the function in the region of the spectrum where the function is defined to be equal to 1. This function is invariable if none of the experimental settings of the spectrometer are changed i.e. identical excitation geometry, identical detector for all samples, identical gain for the amplifier and data gathering electronics, etc. These types of functions are used as a common mechanism to make adjustments and quantification of spectra; for example they are found in the database of the Spectra PICOFOX program, version 7.2.5.0, released by Bruker.

The second vector has information about the sample composition, the physical properties for each element and the detection efficiency for a given energy and the geometric setup. All the parameters, for each element $i$, are defined in Eq.(1).

Once a specimen of the sample is analyzed and the spectrum has been acquired, it can be expressed as:

$$S = I_0 N_0 F_1 (Km\sigma\omega/A)_{z=1} + I_0 N_0 F_2 (Km\sigma\omega/A)_{z=2} + \cdots + I_0 N_0 F_n (Km\sigma\omega/A)_{z=n} \quad (7)$$

Where $n$ is the total number of elements that the technique can determine. Then Eq. (7) can be expressed as:

$$S = \delta_1 F_1 + \delta_2 F_2 + \cdots + \delta_n F_n = \sum_{i=1}^{n} \delta_i F_i \quad (8)$$

Where $\delta_i$ are the characteristic numerical coefficients for each chemical element $i$, which takes into account the composition (or the relative abundance of the elements) of each sample and the intensity of the excitation source.

If we integrate Eq.(8) over all of the studied energies, we obtain:

$$\int_0^{Emax} S\, dE = \int_0^{Emax} (\delta_1 F_1 + \delta_2 F_2 + \cdots + \delta_n F_n)\, dE \quad (9)$$

Since the $\delta_n$ values are numerical coefficients, and all the $Fi$ functions are normalized to have area of 1, the result is:

$$\int_0^{Emax} S\, dE = \int_0^{Emax} \sum_{i=1}^{n} \delta_i F_i\, dE = \sum_{i=1}^{n} \delta_i \left( \int_0^{Emax} F_i\, dE \right) = \sum_{i=1}^{n} \delta_i = \delta_{Total} \quad (10)$$

The value $\delta_{Total}$ can be understood as the total intensity obtained in a given spectrum. So, if we now evaluate the relationship $S/\delta_{Total}$, following the sequence of Eqs.(9) and (10), we obtain:

$$\frac{S}{\delta_{Total}} = \frac{\delta_1 F_1}{\delta_{Total}} + \frac{\delta_2 F_2}{\delta_{Total}} + \cdots + \frac{\delta_n F_n}{\delta_{Total}} = S_N \quad (11)$$

This means that the $S_N$ spectrum has an area of 1.

Now we have to notice that all specimens of the same sample have, for each element $i$, the same value of $\delta_i/\delta_{Total}$, since the composition of the sample remains unchanged in all cases. That means that, if the technique is working in appropriate conditions, the measurement of different specimens of a sample, at different states of dilutions, should produce the same $S_N$ spectrum.

And in case the technique is not working in adequate conditions, the relative changes/differences in the $S_N$ spectra can be quantified.

This type of comparison is valid for any portion of the spectrum. This can be visualized by following the same sequence of Eqs.(7) and (8), and modifying the limits of integration at Eq.(9), reducing them to the Region Of Interest (ROI, between regions $E_1$ and $E_2$ of the spectrum). Later, we follow the sequence of Eqs. (10) and (11), in order to obtain:

$$\int_{E1}^{E2} S_{(ROI)}\, dE = \int_{E1}^{E2} (\delta_1 F_1 + \delta_2 F_2 + \cdots + \delta_n F_n)\, dE = \sum_{i(E1)=1}^{j(E2)} \delta_i = \delta_{(ROI)} \quad (12)$$

$$\frac{S_{(ROI)}}{\delta_{(ROI)}} = \frac{1}{\delta_{(ROI)}} \sum_{i(E1)=1}^{j(E2)} \delta_j F_j = S_{N(ROI)} \quad (13)$$

In order to identify non-linear effects that affect the technique, we can compare either full normalized spectra or arbitrarily defined normalized portions (ROI) of different spectra.

## 3. Methodology.

### 3.1. Instrumental.

In order to obtain comparisons about different TXRF equipment, the same samples were measured in two different laboratories.

Measurements were performed with two bench top TXRF spectrometers: 1) A S2 PicoFox (Bruker AXS, Madison, WI, USA), and 2) A S2 PICOFOX spectrometer (Bruker AXS Microanalysis GmbH, Berlin, Germany).

Both portable instruments are enclosed in an X-ray biological shield, and they are equipped with an air-cooled, low-power, 50 kV X-ray metal-ceramic tube, working at 50 W of maximun power, with a molybdenum target, a multilayer monochromator with 80% reflectivity and a liquid nitrogen-free (Peltier-cooled) XFlash Silicon Drift Detector (SDD) with an energy resolution better than 150 eV (for the Mn K$\alpha$ line, 13.59 keV).

Sample irradiations were done with the X-ray tube at 50 kV, a 1-mA current for irradiation and a data collection time of 500 s.

### 3.2. Materials and Methods.

The calibration of the TXRF spectrometer was performed using eight MERCK ICP single element standard solutions of S, Ti, Cr, Fe, Ni, Zn, Ge and As.

The multielement stock solution (**A**) in a concentration of 1500 mgL$^{-1}$ for K, Fe, Zn, Na; 100 mgL$^{-1}$ for Ga, 40 mgL$^{-1}$ for Ti, Cr, Ni; 10 mgL$^{-1}$ for As and 1 mgL$^{-1}$ for Cu; was prepared from salt dissolution ($KNO_3$ (MallinckrodtTM), $NaNO_3$ (MallinckrodtTM), $FeCl_3$ (MallinckrodtTM), $ZnCO_3$ (MallinckrodtTM)) in deionized water at pH=2.0 adjusted with $HNO_3$ (Anedra ®). All salts were pre-dried at 105˚C for 1 hour. Ti, Cr, Ni, As and Cu from individual commercial stock solutions of 1000 mgL$^{-1}$ (SCP SCIENCE ®) were added.

For sample preparation, pure water was obtained from a feeder 55 WG with subsequent de-ionization by a Milli-Q SP Reagent Water System (Millipore) which yields ultrapure water with specific electrical resistivity higher than 18 MΩ.cm. Nitric acid (Anedra ®) was used for cleaning quartz reflectors and deionized water was employed for rinsing and dilution purposes.

### 3.3. Sample preparation and Measurements.

The solution **A** was sequentially diluted, in 5 stages: **D1** = [A]/3= Co/3; **D2** = (D1)/3; **D3** = (D2)/3; **D4** = (D3)/3 and **D5** = (D4)/3.

Each stage of dilution was measured 3 times, with specimens with 2µL, 3 µL and 5 µL. (18 spectra were acquired, (Co to D5)*3).

## 4. RESULTS.

The 18 spectra were evaluated according to Eqs. (5), (11) and (13), in order to evaluate the non-linear effects in the acquired data.

In every spectrum the intensity for each element (K, Fe, Zn, Ti, Cr, Ni, As, Cr and Ga) was evaluated using the program Spectra PICOFOX version 7.2.5.0 (released by Bruker). Applying the Eq. (5), the relationship for each element compared with the intensity of Ga was pictured as a function of the Dead Time. Figures were divided in 2:, 1) The elements that are manifested with minor intensity and 2) those with the highest intensity in the spectra.

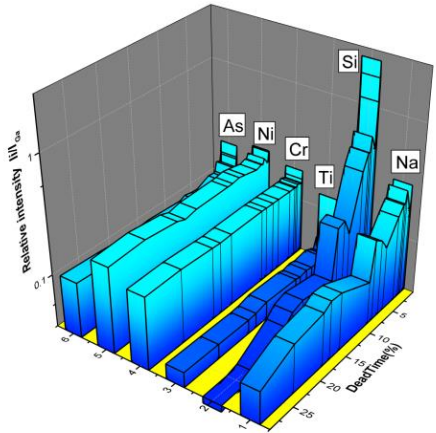

***Figure 2****: Relationship of the intensities of the minor concentration elements, compared with the intensity of Ga, as a function of the Dead Time.*

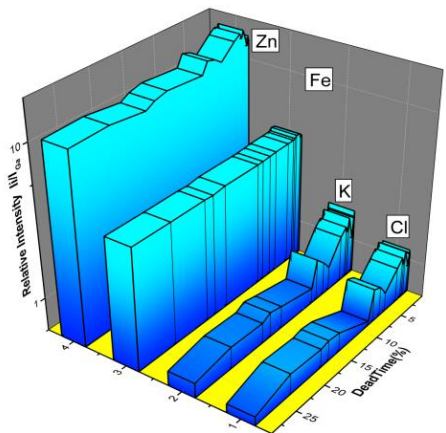

***Figure 3****: Relationship of the intensities of the major concentration elements, compared with the intensity of Ga, as a function of the Dead Time.*

Applying the procedure derived from Eqs. (11) and (13), every spectrum was studied as a set of data ($I_j$ vs $E_j$, counts intensity vs energy, being $j$ the channel number).

We investigate, as a function of the amount of mass in the specimen, the changes in: 1) The relative intensity of the peaks, making comparisons between two normalized spectra, and 2) the shape of each peak, when the spectra were acquired with strong changes in Dead Time. In Fig.(4) two normalized spectra are shown; in black the TXRF normalized spectrum for the original concentrated solution A obtained with a 5 μL specimen. In red is shown the TXRF normalized spectrum for the 27 times dilution of sample A (D3), obtained with a 2 μL specimen.

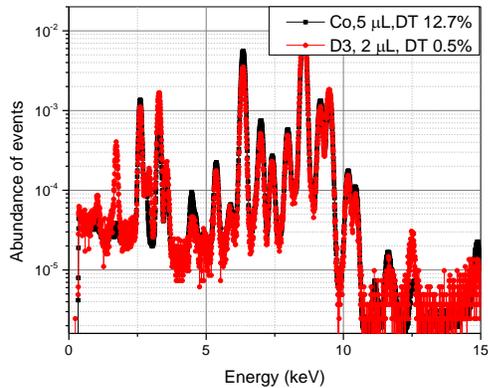

*Figure 4*: The TXRF normalized spectrum obtained with 5 µL specimen of the original solution A, is shown in black. Also, the TXRF normalized spectrum for the 27 times dilution of sample A (D3), obtained with a 2 µL specimen, is shown in red.

In Fig. (4) we clearly notice the difference in between both spectra, which were acquired with significant differences in the amount of sample analyzed. If the technique is properly working, the subtraction between both spectra only should show statistical fluctuations without a systematic tendency, since the analyzed specimens have the same composition. Therefore, the differences observed in Fig. (4) should be exclusively attributed to non-linear effects.

In order to develop a more precise analysis of the non-linear effects, it is convenient to analyze every spectrum in two portions: the first containing low energies (0 to 5 keV), and the second containing medium and high energies (5 to 10 keV.

In Fig. (5) we show the region of low energies (0 to 5 keV) of the spectrum obtained with a high amount of sample (in black, with 12.7% Dead Time). Also we have performed the subtraction between the two spectra shown in Fig. (4), were we obtain as a result how the spectrum shown in black (obtained with high amount of sample) should be modified in order to eliminate the distortions in its shape, owing to non-linear effects. In the portions colored in red this spectrum should be increased; and in blue regions it should be decreased.

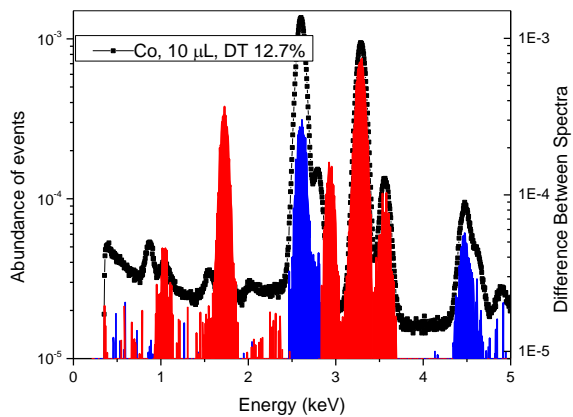

***Figure 5****: Reduced portion of the spectrum (small energies, from 0 to 5 keV). In black is shown the TXRF normalized spectrum obtained with 5 µL of the original solution A. Also it is shown how this spectrum should be modified in order to eliminate the non-linear effects. In red regions, the spectrum should be increased; and in in blue regions it should be decreased.*

The same procedure was made for the region of medium energies (5 to 10 keV). The results are shown in Fig. (6).

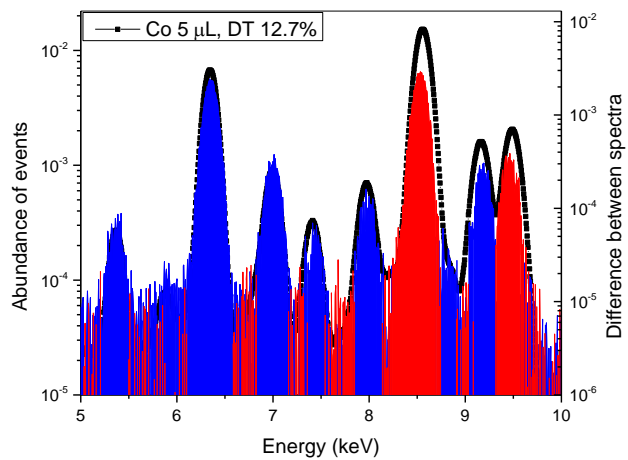

***Figure 6****: Reduced portion of the spectrum (medium energies, from 5 to 10 keV). In black is shown the TXRF normalized spectrum obtained with 5 µL of the original solution A. Also it is shown how this spectrum should be modified in order to eliminate the non-linear effects. In red regions, the spectrum should be increased; and in in blue regions it should be decreased.*

In order to determine if the non-linear effects are more suitable to be studied with another kind of programs, instead those commonly used in TXRF spectroscopy, we have reduced the region of interest in the spectrum to the energies from 8.2 to 8.9 keV, where the most important normalized peak in the studied sample (*Zn,* K$\alpha$ line) is located. In Fig. 7 are shown simultaneously the TXRF normalized spectrum obtained with 5 µL of the original solution A, in black; and the spectrum obtained with 2 µL of the sample obtained from 27 times dilution of the original sample A, in red. Also the differences between them are shown in blue, which are very small. This negligible difference indicates that the study of TXRF spectra affected by Dead Time with another family of programs, is unnecessary.

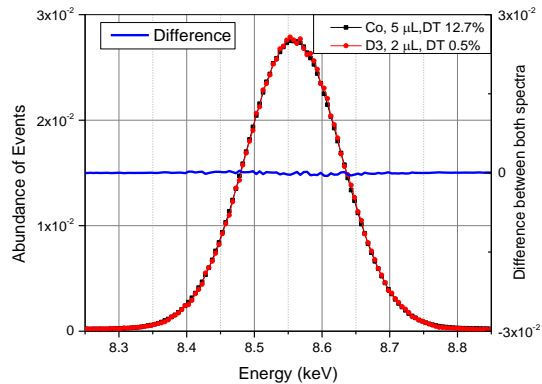

***Figure 7***: *Reduced portion of the spectrum, from 8.2 to 8.9 keV (Zn, K$\alpha$ line, single peak) where in black is the TXRF normalized spectrum obtained with 5 µL of the original solution A. Also, the TXRF normalized spectrum for the 27 times dilution of the original sample, obtained with a 2 µL specimen, is shown in red. The difference between both peaks is shown in blue.*

This same evaluation can be made for each one of the dominant peaks of the spectrum, but in all cases the same result is found. So the non-linear effects do not distort the shape of each peak, instead they produce alterations in the relative abundance of the acquired peaks. That means that the use of a new generation of programs would be irrelevant.

## Conclusions.

In this work we made a pragmatic evaluation of non-linear effects in TXRF spectroscopy studying the influence of the amount of sample analyzed in order to evaluate physical and technological aspects of the technique. Until now the nonlinear aspects were studied only taking into account the physical interactions in the specimen (matrix effects). Here we included the technological limitations of the technique (related with the acquisition and processing electric signals in the detection chain) which showed to be more important, for the sort of samples studied here.

This is essentially a data analysis job which can be divided in 3 main activities: 1) The measuring of the samples, 2) data analysis and interpretation, and 3) the application of the conclusions obtained, for the development of programs to be applied to measurements affected by non-linear effects, in order to improve the reliability of the results, if it would be possible.

The experimental evaluations of the same multi-element solution were made at five different dilution stages. So the TXRF spectra were obtained with high, medium and low concentrations. In order to generalize the results the measurements were replicated by 2 independent laboratories. The inorganic samples studied, and the $SiO_2$ sample holders used, produce a much lower background than the organic samples or the polyethylene sample holders, so they are a particular kind of samples which are less affected by Dead Time.

Concerning the data analysis and interpretation, we have evaluated two theoretical formulations. In both, the changes in the relative intensity of the peaks and the shape of the peaks are studied as a function of the amount of sample. According to Figs. (2), (3), (5) and (6), we observe that: 1) For the type of inorganic samples studied, the Dead Time effect is more important than the Matrix Effects. At medium energies of the spectrum it is observed that the peaks of higher energy are, in general, reduced and those of lower energy are increased. 2) Each sample has its own function that eliminates the undesired effects of Dead Time, which depends on the relative abundance (concentration) of the elements in the sample. Therefore, there would be no general formula to remove or restore the unwanted/undesired effects of Dead Time on acquired spectra. On the other hand, at low energies some peaks tend to disappear as a function of Dead Time, so it is extremely difficult to propose a method that allows to rescue the information. 3) According to Fig. (7) the Dead Time almost does not distort the shape of the individual peaks, it only modifies their intensity, at least with the two Bruker equipment used in this study. Therefore the non-linear effects do not distort the shape of each peak, instead they produce alterations in the relative abundance of the acquired peaks. That means that the use of a new generation of programs cannot help this particular issue.

Taking into account the conclusions of this work: 1) Some usually accepted characteristics of the technique could be re-evaluated, like the 5 orders of magnitude of dynamic range, or the limits of percentages measured in solids; 2) The same study, following the same sequence, could be carried out with other equipments in order to determine the characteristic response of each one; 3) It may be convenient to report in TXRF results the Dead Time values, in addition to the respective concentrations obtained; 4) In low Z samples the background is greatly increased, so in specialized areas with biological samples such as medical, metallomics, biopharmaceutical, clinical research, etc. a high Dead Time is produced. In order to obtain reliable results, it would be recommended to control the values of Dead Time in TXRF measurements in these fields; 5) The non-linear effects could be responsible for systematic differences found in some inter-comparisons between TXRF with other techniques.[14,15,16]

Finally, since there is a strong variation in the relative intensity of the Silicon peak as a function of the Dead Time (as is shown in Appendix 1), to make Silicon determinations using the subtraction technique would not be possible. So, for TXRF Silicon determinations, it is highly recommended to use Polyetilenne (or silicon free materials) as sample holders.

## Acknowledgements

L.B. thanks to H.B. for helpful discussions and a R.V. for samples preparation.

## Appendix 1. Relative intensity of the Silicon peak as a function of the Dead Time.

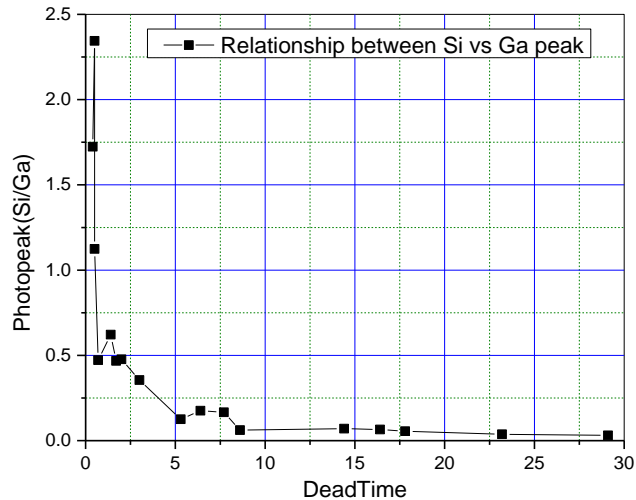

*Figure A1.1. Relationship between the Silicon and the Gallium intensities, as a function of the Dead Time (%).*

It is not feasible to calculate the Silicon peak´s intensity by means of a difference in measurements, from the sample holder minus the measurement of the sample with Silicon, since there is a strong variation of the Silicon signal as a function of the Dead Time.

## References.